\documentclass[twocolumn]{revtex4-1} 
\usepackage{amsmath}
\usepackage{amssymb}
\usepackage{minibox}
\usepackage{microtype}
\usepackage{etoolbox}
\usepackage{bm}
\usepackage[applemac]{inputenc}
\usepackage[usenames,dvipsnames,svgnames,table]{xcolor}
\usepackage{graphicx}

\begingroup
\renewcommand*{\arraystretch}{1.5}
\endgroup
\makeatletter
\renewcommand*\env@matrix[1][\arraystretch]{%
  \edef\arraystretch{#1}%
  \hskip -\arraycolsep
  \let\@ifnextchar\new@ifnextchar
  \array{*\c@MaxMatrixCols c}}
\makeatother

\makeatletter
\apptocmd{\thebibliography}{\global\c@NAT@ctr 10\relax}{}{}
\makeatother

\begin{document}

\title{Observation of chiral edge states with neutral fermions in synthetic Hall ribbons} 

\author{
M. Mancini$^{1}$, G. Pagano$^{1}$, G. Cappellini$^{2}$, L. Livi$^{2}$, M. Rider$^{5,6}$\\J. Catani$^{3,2}$, C. Sias$^{3,2}$, P. Zoller$^{5,6}$, M. Inguscio$^{4,1,2}$, M. Dalmonte$^{5,6}$, L. Fallani$^{1,2}$
}

\affiliation{
$^1$\minibox{Department of Physics and Astronomy, University of Florence, 50019 Sesto Fiorentino, Italy}\\
$^2$\minibox{LENS European Laboratory for Nonlinear Spectroscopy, 50019 Sesto Fiorentino, Italy}\\
$^3$\minibox{INO-CNR Istituto Nazionale di Ottica del CNR, Sezione di Sesto Fiorentino, 50019 Sesto Fiorentino, Italy}\\
$^4$\minibox{INRIM Istituto Nazionale di Ricerca Metrologica, 10135 Torino, Italy}\\
$^5$\minibox{Institute for Quantum Optics and Quantum Information of the Austrian Academy of Sciences, A-6020 Innsbruck, Austria}
$^6$\minibox{Institute for Theoretical Physics, University of Innsbruck, A-6020 Innsbruck, Austria}
}

\begin{abstract}
Chiral edge states are a hallmark of quantum Hall physics. In electronic systems, they appear as a macroscopic consequence of the cyclotron orbits induced by a magnetic field, which are naturally truncated at the physical boundary of the sample. Here we report on the experimental realization of chiral edge states in a ribbon geometry with an ultracold gas of neutral fermions subjected to an artificial gauge field. By imaging individual sites along a synthetic dimension, we detect the existence of the edge states, investigate the onset of chirality as a function of the bulk-edge coupling, and observe the edge-cyclotron orbits induced during a quench dynamics. The realization of fermionic chiral edge states is a fundamental achievement, which opens the door towards experiments including edge state interferometry and the study of non-Abelian anyons in atomic systems.
\end{abstract}

\maketitle

Ultracold atoms in optical lattices represent an ideal platform to investigate the physics of condensed-matter problems in a fully tunable, controllable environment \cite{lewenstein2012,bloch2012}. One of the remarkable achievements in recent years has been the realization of synthetic background gauge fields, akin to magnetic fields in electronic systems. Indeed, by exploiting light-matter interaction, it is possible to imprint a Peierls phase onto the atomic wavefunction, which is analogous to the Aharanov-Bohm phase experienced by a charged particle in a magnetic field \cite{jaksch2003,dalibard2011,goldman2014}. These gauge fields, first synthesized in Bose-Einstein condensates \cite{lin2009}, have recently allowed for the realization of the Harper-Hofstadter Hamiltonian in ultracold bosonic 2D lattice gases \cite{aidelsburger2013,miyake2013}, paving the way towards the investigation of different forms of bulk topological matter in bosonic atomic systems \cite{goldman2014,duca2015}. In the present work we are instead interested in the edge properties of fermionic systems under the effects of a synthetic gauge field. Fermionic edge states are a fundamental feature of 2D topological states of matter, such as quantum Hall and chiral spin liquids~\cite{wenbook,girvinbook}. Moreover, they are robust against changing the geometry of the system by keeping its topology, and can be observed even on Hall ribbons~\cite{celi2014}. In addition, they offer very attractive perspectives in quantum science, such as the realization of robust quantum information buses~\cite{yao2013}, and they are ideal starting points for the realization of non-Abelian anyons akin to Majorana fermions~\cite{nayak2008,lindner2012}.

\begin{figure}[t!]
\begin{center}
\includegraphics[width=\columnwidth]{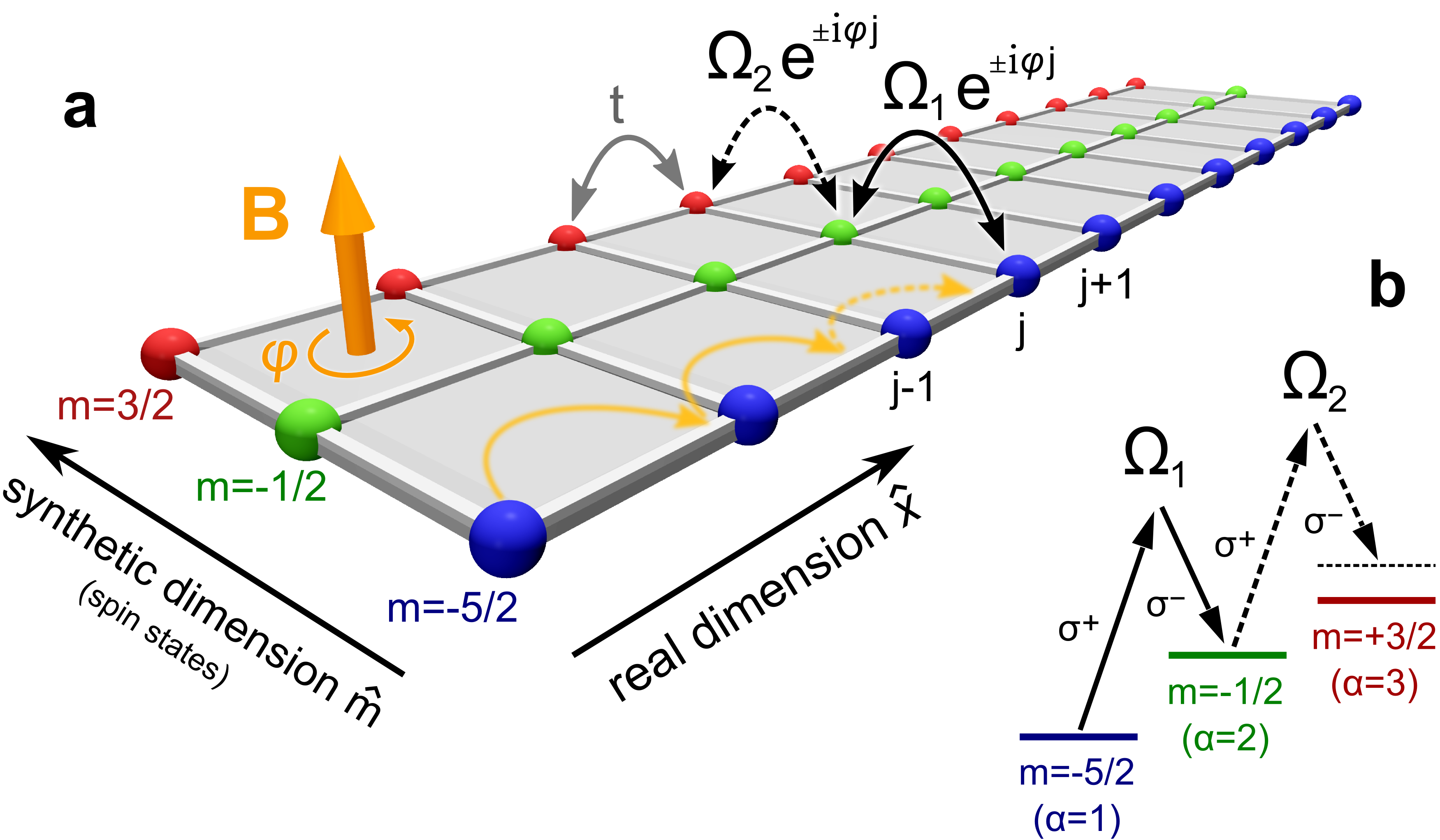}
\end{center}
\caption{{\bf A synthetic gauge field in a synthetic dimension.} {\bf a.} We confine the motion of fermionic ultracold atoms in a hybrid lattice, generated by an optical lattice along a real direction $\hat{x}$ with tunneling $t$, and by a laser-induced hopping between spin states along a synthetic direction $\hat{m}$. By inducing a complex tunneling $\Omega_{1,2} e^{i\varphi j}$ along $\hat{m}$, the atom wavefunction acquires a phase $\varphi$ per plaquette, mimicking the effect of a transverse magnetic field $\mathbf{B}$ on effectively charged particles. {\bf b.} Scheme of the $^{173}$Yb nuclear spin states and Raman transitions used in the experiment.} \label{fig:fig1}
\end{figure}

Here, we report the observation of chiral edge states in a system of neutral fermions subjected to a synthetic magnetic field. We exploit the high level of control in our system to investigate the emergence of chirality as a function of the Hamiltonian couplings. These results have been enabled by an innovative experimental approach, where an internal (nuclear spin) degree of freedom of the atoms is used to encode a lattice structure lying in an ``extra dimension''~\cite{celi2014}, providing direct access to edge physics. In addition, we validate the chiral nature of our system by performing quench dynamics, demonstrating how the particle motion
shows edge-cyclotron orbits \cite{note1}. 

\begin{figure*}[t!]
\begin{center}
\includegraphics[width=0.9\textwidth]{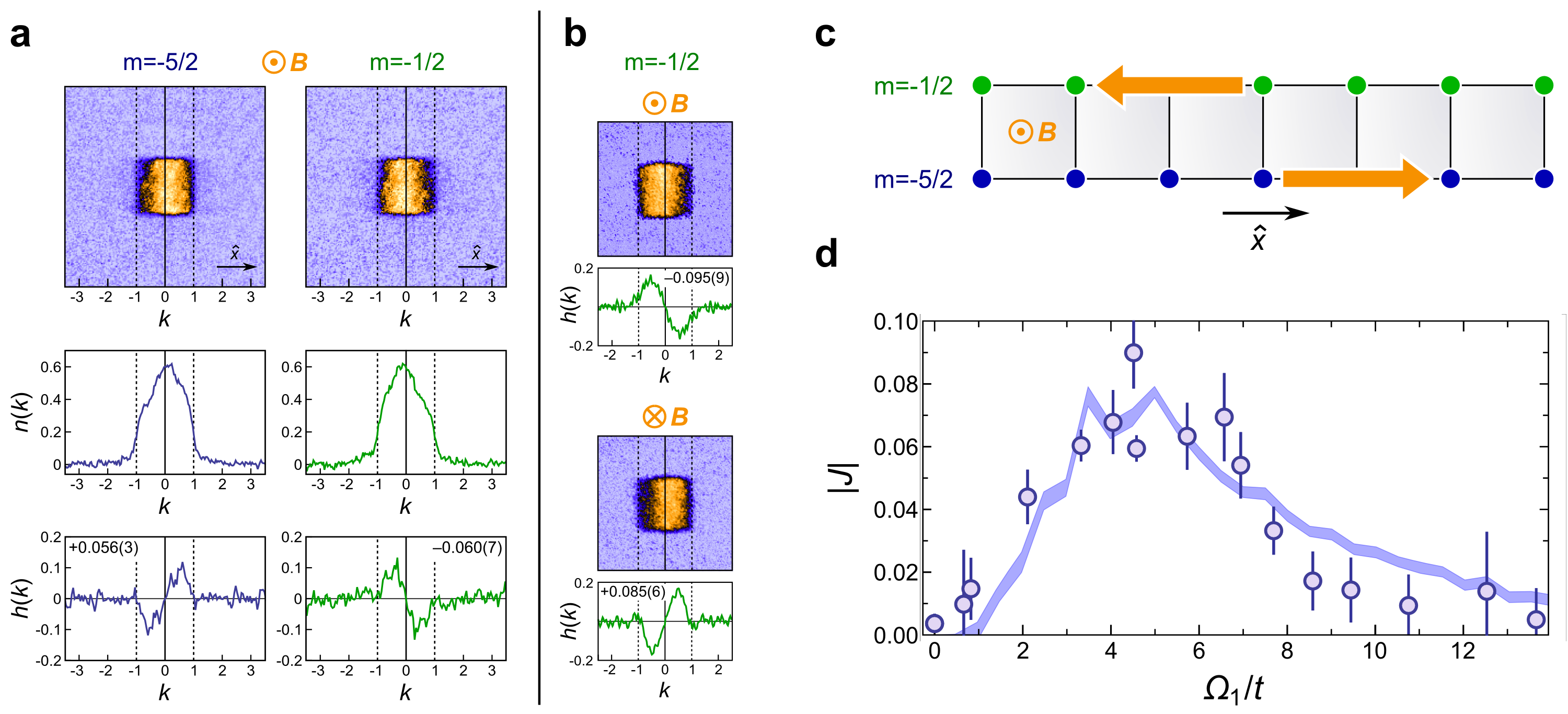}
\end{center}
\caption{
{\bf Chiral dynamics in 2-leg ladders.} {\bf a.} The upper panel shows false-color time-of-flight images of the atoms in the $m=-5/2$ and $m=-1/2$ legs (averages of $\sim 30$ realizations). The central panels show the integrated lattice momentum distribution $n(k)$ and the lower panels show $h(k)$ (the numbers in the insets are the values of $J$ determined from $h(k)$). Experimental parameters: $\Omega_1=2\pi \times 489$ Hz, $t=2\pi \times 134$ Hz, $\Omega_1/t=3.65$, $\varphi=0.37\pi$. {\bf b.}  Time-of-flight images and $h(k)$ of the $m=-1/2$ leg for opposite values of the effective magnetic field. Experimental parameters: $\Omega_1=2\pi \times 394$ Hz, $t=2\pi \times 87$ Hz, $\Omega_1/t=4.53$, $\varphi=\pm0.37\pi$. {\bf c.} Sketch of the 2-leg ladder configuration realized for this experiment. The arrows are a pictorial representation of the chiral currents.
{\bf d.} The circles show experimental values of $|J|$ for the $m=-1/2$ leg as a function of $\Omega_1/t$ (averages of datasets taken for $\varphi=\pm0.37\pi$). The error bars are obtained with a bootstrapping method applied on $\sim 30$ different measurements. The shaded area is the result of a numerical simulation including thermal fluctuations (see Supplementary Materials).
} \label{fig:fig2}
\end{figure*}

We synthesize a system of fermionic particles in an atomic Hall ribbon of tunable width pierced by an effective gauge field. One dimension is realized by an optical lattice, which induces a {\it real} tunneling $t$ between different sites along direction $\hat{x}$ (see Fig. 1a).
The different internal spin states are coupled by a two-photon Raman transition, which provides a coherent controllable coupling $\Omega e^{i \varphi x}$ between different spin components. This can be interpreted as a complex tunneling amplitude between adjacent sites of an "extra-dimensional" lattice \cite{boada2012,celi2014}. Furthermore, the phase imprinting laid out by the Raman beams amounts to the synthesis of an effective magnetic field for effectively charged particles \cite{dalibard2011} with flux $\varphi/2\pi$ (in units of the magnetic flux quantum) per plaquette (see Supplementary Materials). The Hamiltonian describing the system is
\begin{eqnarray}
H&=& \sum_{j}\sum_{\alpha}\left[-t(c^{\dagger}_{j, \alpha}c_{j+1, \alpha}+\textrm{h.c.}) + \mu_j n_{j, \alpha}\right]\nonumber\\
&+& \sum_{j}\sum_{\alpha}\left[ \frac{\Omega_\alpha}{2}(e^{i\varphi j}c^\dagger_{j,\alpha}c_{j,\alpha+1}+\textrm{h.c.}) + \xi_\alpha n_{j,\alpha}\right]
\end{eqnarray}
where $c^\dagger_{j,\alpha}(c_{j,\alpha})$ are fermionic creation (annihilation) operators on the site $(j,\alpha)$ in the real ($j$) and synthetic ($\alpha = 1, 2, 3$) dimension, and $n=c^\dagger_{j,\alpha}c_{j,\alpha}$. The first line describes the dynamics along $\hat{x}$, where $t$ can be tuned by changing the intensity of the optical lattice beams. The dynamics along $\hat{m}$ is encoded in the second line: $\Omega_\alpha$ can be controlled by changing the power of the Raman beams, while $\varphi$ can be tuned by changing their angle (see Supplementary Materials). Besides the tunneling terms, $\mu_j$ describes a weak trapping potential along $\hat{x}$, while $\xi_\alpha$ accounts for a state-dependent light shift, providing an energy offset along $\hat{m}$. In our experiment we realize large synthetic magnetic fields corresponding to $\varphi \simeq 0.37 \pi$ per plaquette. The carriers are constituted by alkaline-earth-like $^{173}$Yb atoms, initially prepared in a degenerate Fermi gas, and the sites of the synthetic dimension (see Fig. 1b) are encoded in a subset of spin states $\left\{ m \right\}$ out of the $I=5/2$ nuclear spin manifold, thus providing up to 6-leg ladders. These atoms show SU($N$)-invariant interactions \cite{cazalilla2014}, inhibiting the redistribution of the atoms among the different synthetic sites by collisional processes \cite{taie2012,pagano2014}.

\begin{figure*}[t!]
\begin{center}
\includegraphics[width=0.9\textwidth]{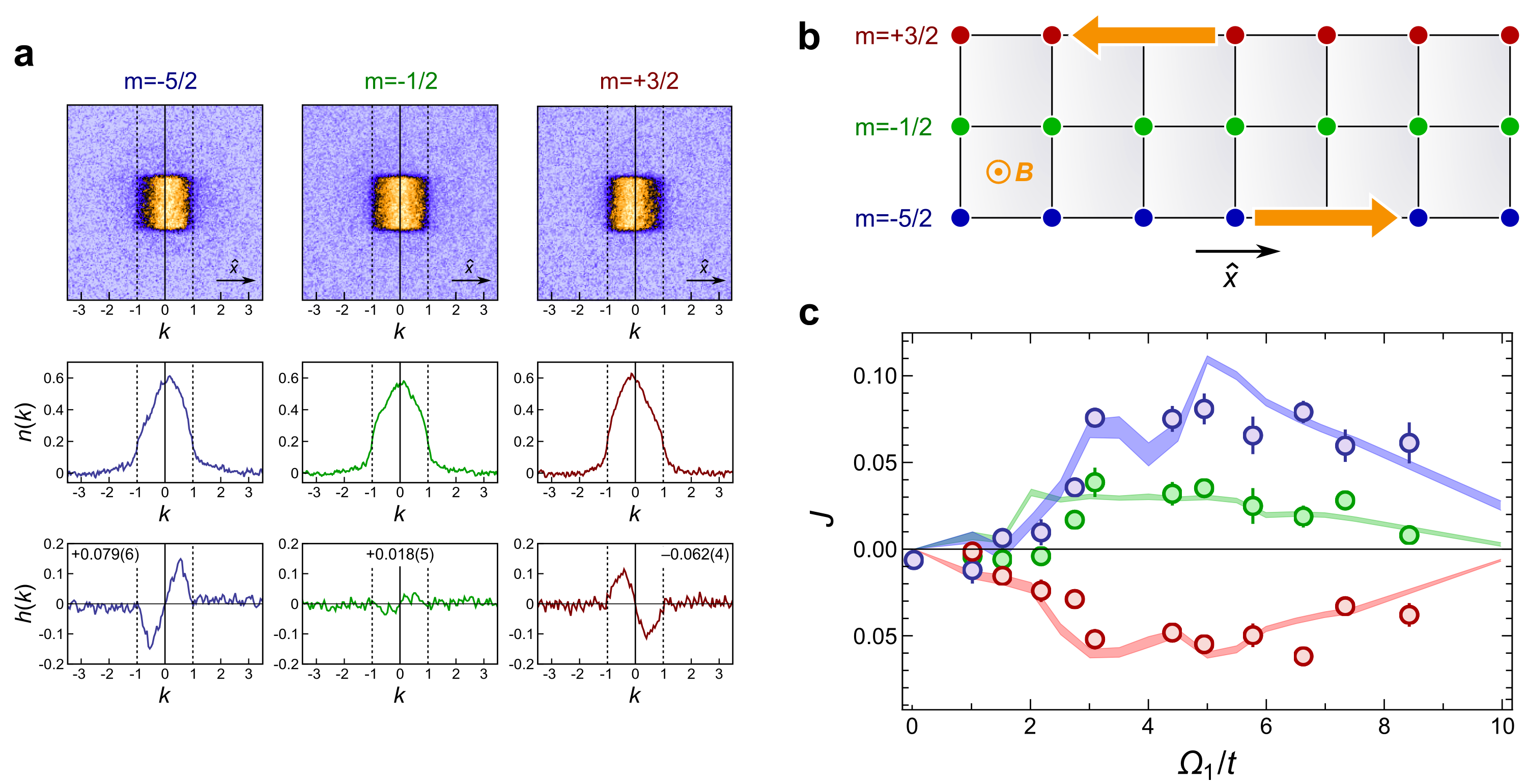}
\end{center}
\caption{{\bf Chiral edge currents in a 3-leg ladder.} {\bf a.} Experimental time-of-flight images (up), $n(k)$ (center) and $h(k)$ for each of the three legs $m=-5/2$, $m=-1/2$ and $m=+3/2$ constituting the ladder (the numbers in the insets of the graphs are the values of $J$ determined from $h(k)$). Experimental parameters: $\Omega_1=2\pi \times 620$ Hz, $t=2\pi \times 94$ Hz, $\Omega_1/t=6.60$, $\varphi=0.37\pi$. {\bf b.} Sketch of the 3-leg ladder configuration realized for this experiment. {\bf c.} The circles show experimental values of $J$ for each leg as a function of $\Omega_1/t$. The shaded areas are the results of a numerical simulation (see Supplementary Materials). For both experimental and simulation data, the blue, green, and red color correspond to $m=-5/2$, $m=-1/2$, and $m=+3/2$ respectively.} \label{fig:fig3}
\end{figure*}

The key advantage offered by the implementation of the lattice in a real+synthetic space is the possibility to work with a finite-sized system with {\it sharp} and {\it addressable} edges. Specifically, we focus on elementary configurations made up of fermionic ladders with a small number of ``legs'' connected by a tunnel coupling between them. A leg is constituted by a 1D chain of atoms trapped in the sites of the real lattice in a specific spin state, whereas the ``rungs" are provided by the synthetic tunneling (see Fig. 1a). The number of legs can be set by controlling the light shifts $\xi_\alpha$ in such a way as to choose the number of spin states that are coupled by the Raman lasers (see Supplementary Materials).

We first consider the case of a 2-leg ladder constituted by the nuclear spin states $m=-5/2$ and $m=-1/2$. A quantum degenerate $^{173}$Yb Fermi gas, at an initial temperature $T\simeq 0.2T_F$ (where $T_F$ is the Fermi temperature), is first spin-polarized in $m=-5/2$. By slowly turning on the intensity of the real-space lattice, we prepare a system of ladders where all the atoms occupy the $m=-5/2$ leg with less than one atom per site (i.e. in a conductive metallic state). Then, by controlling the intensity and frequency of the Raman beams (see Supplementary Materials), we slowly activate the tunnel coupling between the legs in such a way as to load the Hall ribbons in an equilibrium state, without populating excited bands. 

\begin{figure*}[t!]
\begin{center}
\includegraphics[width=0.75\textwidth]{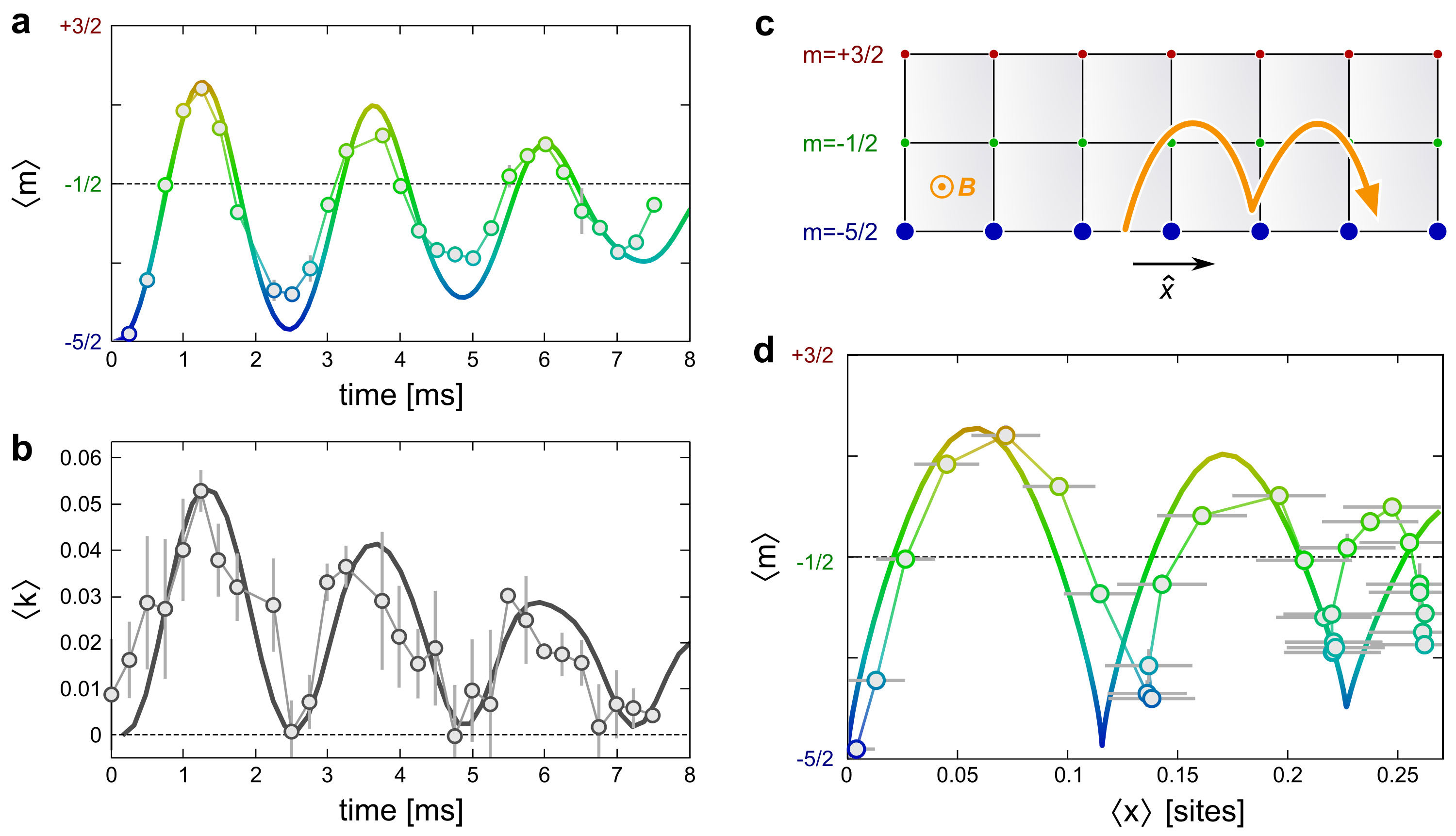}
\end{center}
\caption{{\bf Edge-cyclotron orbits.} {\bf a.} Time dependence of the average position in the synthetic direction $\langle m \rangle$ after a quench on the synthetic tunneling. {\bf b.} Time dependence of the average lattice momentum $\langle k \rangle$ along the $\hat{x}$ direction. {\bf c.} Schematics of the edge-cyclotron orbits. {\bf d.} Average position in $\hat{m}-\hat{x}$ space. The circles in panels {\bf a}, {\bf b}, {\bf d} are experimental data, the thin lines connect the points and the thick lines are the theoretical predictions. Experimental parameters: $\Omega_1=2\pi \times 490$ Hz, $t=2\pi \times 94$ Hz. After the second orbit in panel {\bf d}, the mismatch between theory and experiment could be possibly ascribed to an accumulation of integration error in the data analysis, which amplifies the effects of the assumptions in the model (not including, for instance, interactions).} \label{fig:fig4}
\end{figure*}

Despite the absence of a real bulk region, this 2-leg configuration is expected to support chiral currents with atoms flowing in opposite directions along the legs (see Fig. 2c), as investigated recently in bosonic systems \cite{atala2014}.
In order to witness this property, we measure the relative motion of the atoms in the two legs by spin-selective imaging of the lattice momentum distribution, obtained switching off the synthetic coupling and releasing the atoms from the lattice.
In Fig. 2a (upper panel) we show two time-of-flight images referring to both the $m=-5/2$ and the $m=-1/2$ legs (see Fig. 2c) for $\Omega_1 = 2\pi \times 489$ Hz and $t=2\pi \times 134$ Hz ($\Omega_1/t=3.65$). Here we are interested only in direction $\hat{x}$, which reflects the distribution of the lattice momenta $k$ along the legs (in units of the real-lattice wavenumber $k_L=\pi/d$, where $d$ is the real-lattice spacing). The lattice momentum distribution in the transverse direction is a uniform square due to the presence of an additional optical lattice, required to suppress the dynamics along the $\hat{y}$ and $\hat{z}$ real directions (see Supplementary Materials). The central panel of Fig. 2a shows the lattice momentum distribution $n(k)$ after integration of the images along the transverse direction and normalization according to $\int n(k) dk =1$. We observe a clear asymmetry in $n(k)$ (similarly to what reported in experiments with spin-orbit coupling in harmonically trapped gases \cite{lin2011,wang2012,cheuk2012}), which we characterize by defining the function 
\begin{equation}
h(k)=n(k)-n(-k) \; , 
\end{equation}
reported in the lower panel of Fig. 2a. The quantity $J = \int_0^1 h(k) dk $ provides a measurement of the lattice momentum unbalance and quantifies the strength of the chiral motion of the particles along the two legs. The values $J=+0.056(3)$ for $m=-5/2$ and $J=-0.060(7)$ for $m=-1/2$ are approximately equal in intensity and opposite in sign, directly evidencing the presence of chirality in the system. We also perform the same experiment with a reversed direction of the synthetic magnetic field $\mathbf{B}$ (see Fig. 2b), observing a change of sign in $J$, corresponding to currents circulating in the opposite direction. This behavior confirms the interpretation of our data in terms of chiral currents induced by a synthetic magnetic field in a synthetic 2D lattice.

The stability of chiral edge states in fermionic systems is of key importance, e.g. for quantum information applications. In our system, the appearance of a chiral behavior is governed by several key parameters, including the ratio $\Omega_1/t$, the Fermi energy $E_F$ and the flux $\varphi$. Their ample tunability can be used to investigate the rise and fall of the edge currents as a function of the Hamiltonian parameters \cite{atala2014}, and to identify which regimes exhibit stronger chiral features. In particular, by varying the tunneling rates along $\hat{x}$ and $\hat{m}$, we observe a phase transition between a chiral behavior and a non-chiral regime. The lattice momentum distribution is measured as a function of $\Omega_1/t$ without affecting other relevant parameters, such as $E_F$ and $T$. Fig. 2d reports the measurement of $|J|$ as a function of $\Omega_1/t$ (circles). As expected, no chirality is observed for vanishing $\Omega_1$, when the legs are decoupled. Interestingly, chirality is also suppressed for large inter-edge coupling $\Omega_1\gg t$. In the latter regime, the largest energy scale in the system is the effective kinetic energy along the synthetic direction: this contribution is minimized when the fermions occupy the lowest energy state on each rung, which does not exhibit any chiral behavior. The measured values of $|J|$ compare well with the results of a numerical simulation including thermal fluctuations (shaded area), as described in the Supplementary Materials.

We have then considered a 3-leg ladder, which is the minimal configuration where chiral currents at the edges can be sharply distinguished from the behavior of the bulk.
The experimental procedure is analogous to that employed for the 2-leg case, adjusting now the Raman parameters in such a way as to extend the synthetic coupling to $m=+3/2$, with $\Omega_2 \simeq 1.41 \Omega_1$ (see Supplementary Materials). Fig. 3a shows experimental data of $n(k)$ and $h(k)$ for each of the three legs for $\Omega_1=2\pi \times 620$ Hz and $t=2\pi \times 94$ Hz ($\Omega_1/t=6.60$). We observe strong chiral currents in the upper and lower edge chains, showing values of $J$ with opposite sign as in the 2-leg case ($J=+0.079(6)$ for $m=-5/2$ and $J=-0.062(4)$ for $m=+3/2$). The central leg, instead, shows a much reduced asymmetry in $n(k)$ ($J=0.018(5)$), signaling a suppressed net current in the bulk. This is a direct evidence of the existence of chiral states propagating along the edges of the system, which leave the bulk mostly decoupled from the edges (see Fig. 3c).
This behavior is akin to what is expected for a fermionic system in a Harper-Hofstadter Hamiltonian. Bulk states exhibit only {\it local} circulations of current, which average to zero when all the different states enclosed by the Fermi surface are considered. Only the edges of the system experience a nonzero current, since there the chiral nature of the states prevents that cancellation effect to occur. We note that in the ribbon geometry of the experiment the bulk reduces to just a single central line. Nevertheless, the behavior discussed above is clearly present and detectable in the experimental signal. Actually, the small width of the ribbon favours the observation of edge states, given the large boundary/surface ratio of the system, which is reflected in a substantial population of states with edge character.

Fig. 3c shows the values of $J$ for the three different legs of the ladder as a function of $\Omega_1/t$.  The results illustrate the role of the bulk-edge coupling: similarly to the 2-leg case, for small coupling chirality is very weak, and increases as $\Omega_1/t$ approaches $\sim 3$. The theoretical curves show that further increasing $\Omega_1/t$ eventually leads to an attenuation of the signal because of the effective coupling between the edges, which smoothens the chiral features of the system. We observe a substantial agreement between experiment and theory for the range of $\Omega_1/t$ that can be explored in our setup. The nonzero current in the bulk can be ascribed to the different couplings $\Omega_1$ and $\Omega_2$, and to a residual light shift which breaks the symmetry between the two edges (see Supplementary Materials).

Finally, we have performed additional quench dynamics experiments that directly evidence the properties of chiral transport along the edges. A system of lattice fermions is prepared in an initial state with zero average momentum on the lower $m=-5/2$ leg of a 3-leg ladder. We then perform a quench by suddenly activating the complex tunneling in the synthetic direction. Fig. 4a shows the time dependence of the average position in the synthetic direction $\langle m \rangle$, measured by optical Stern Gerlach detection~\cite{pagano2014}. Fig. 4b shows the time dependence of the average lattice momentum $\langle k \rangle$ along $\hat{x}$, measured by time-of-flight imaging of the whole cloud. Fig. 4d shows an experimental reconstruction of the average orbit on the ribbon surface as a plot of $\langle m \rangle$ vs. the average position in real space $\langle x \rangle$. The latter has been determined by evaluating the average velocity along $\hat{x}$ from the knowledge of the energy band dispersion vs. lattice momentum, and then performing an integration in time (see Supplementary Materials). The experimental data are in very good agreement with the theoretical predictions, shown by the lines. The dynamics displays a strong chiral character, demonstrated by the in-phase oscillations in Figs. 4a-b and by the orbits in Fig. 4d. Under the effect of the synthetic magnetic field the fermions move according to cyclotron-type dynamics, which is however naturally truncated by the synthetic edge, giving rise to a skipping-type orbit \cite{wenbook,girvinbook}. This dynamics is effectively damped, already at the theoretical level (see Figs. 4a-b), as a result of the average over many different fermionic trajectories, which also determines a reduction of the average orbit radius to less than one lattice site (see Fig. 4d). This is remarkably different from the behavior of a non-interacting Bose gas, which would occupy a single condensed wavepacket undergoing undamped oscillations.

In conclusion, we have reported the existence of chiral edge states with neutral fermions in a quantum Hall ribbon pierced by an artificial gauge field. Our approach can be extended to wide ladder systems with up to $2I+1$ legs, providing a setting for the investigation of both edge and bulk 2D topological matter complementary to recent works on Chern insulators \cite{jotzu2014}. This would allow the study of the combined effect of interactions and synthetic gauge fields, a fundamental ingredient for fractional quantum Hall physics, in a controlled manner, 
potentially leading to the realization of novel states of matter in ladder systems such as, e.g., chiral Mott insulator states. Moreover, the flexibility offered by the present scheme allows engineering arbitrary lattice patterns in ladder systems, including disorder and constriction. This opens the door towards the realization of interferometers for chiral liquids, investigating their transport properties, and the possibility of implementing interfaces between chiral edges potentially hosting exotic non-Abelian anyons such as Majorana-like states \cite{nayak2008}. 

{\bf Acknowledgements.} We would like to thank A. Celi and P. Massignan for early stimulating discussions on the ''synthetic dimension`` approach. M.D. and M.R. thanks C. Laflamme and A. Sterdyniak for discussions. The experimental work in Florence has been supported by EU FP7 SIQS, MIUR PRIN2012 AQUASIM, ERC Advanced Grant DISQUA. The theoretical work in Innsbruck has been supported by ERC Synergy grant UQUAM, SFB FoQuS of the Austrian Science Fund, and EU FP7 SIQS.

\renewcommand{\thefigure}{S\arabic{figure}}
 \setcounter{figure}{0}
\renewcommand{\theequation}{S.\arabic{equation}}
 \setcounter{equation}{0}
 \renewcommand{\thesection}{S.\Roman{section}}
\setcounter{section}{0}
\renewcommand{\thetable}{S\arabic{table}}
 \setcounter{table}{0}

\onecolumngrid

\newpage


\begin{center}
{\bf \large Supplementary Materials for\\
``Observation of chiral edge states with neutral fermions in synthetic Hall ribbons''}

\bigskip

M. Mancini, G. Pagano, G. Cappellini, L. Livi, M. Rider\\J. Catani, C. Sias, P. Zoller, M. Inguscio, M. Dalmonte, L. Fallani

\end{center}

\bigskip
\twocolumngrid

\section{Experimental setup}

The starting point for our experiment is a $^{173}$Yb spin-polarized degenerate Fermi gas with $N_{\mathrm{at}}\simeq 1.6\times 10^4$ atoms at a temperature $T \simeq 0.2T_F$ (where $T_F$ is the Fermi temperature). Quantum degeneracy is achieved by forced evaporation of a ($m=-5/2$) + ($m=+5/2$) nuclear spin mixture in an optical dipole trap with mean geometric frequency $\bar{\omega}/2\pi\simeq 80~\mbox{Hz}$. After evaporation, the atoms in the $m=+5/2$ state are removed by a resonant laser pulse and a spin-polarized Fermi gas in the $m=-5/2$ state is left. The atoms are then confined in a three-dimensional cubic optical lattice with periodicity $d=\lambda_L/2=380~\mbox{nm}$. The three lattice depths are set to $V_{0x}=6.5 E_R$ and $V_{0y}=V_{0z}=30 E_R$  (where $E_R=h^2/2M\lambda_L^2$ is the recoil energy, $h$ is the Planck's constant and $M$ is the atomic mass). Along $\hat{y}$ and $\hat{z}$ the tunneling rates ($t_{y,z} / 2\pi \simeq 1~\mbox{Hz}$) are negligible on the timescale of the experiment, leading to the realization of an array of  $\approx 1000$ independent 1D fermionic chains characterized by a longitudinal harmonic confinement with frequency $\omega_x/2\pi \simeq 55$ Hz. The dynamics is allowed only in the shallow lattice along the $\hat{x}$ direction where $t/2\pi \sim 90~\mbox{Hz}$. The lattice occupation is less than 1 atom per site on the central chain, ensuring that the atoms are in a metallic state in the lowest energy band.

\section{Realization of the fermionic Hall ribbons}

\begin{figure}[t!]
\centering
\includegraphics[width=0.72\columnwidth]{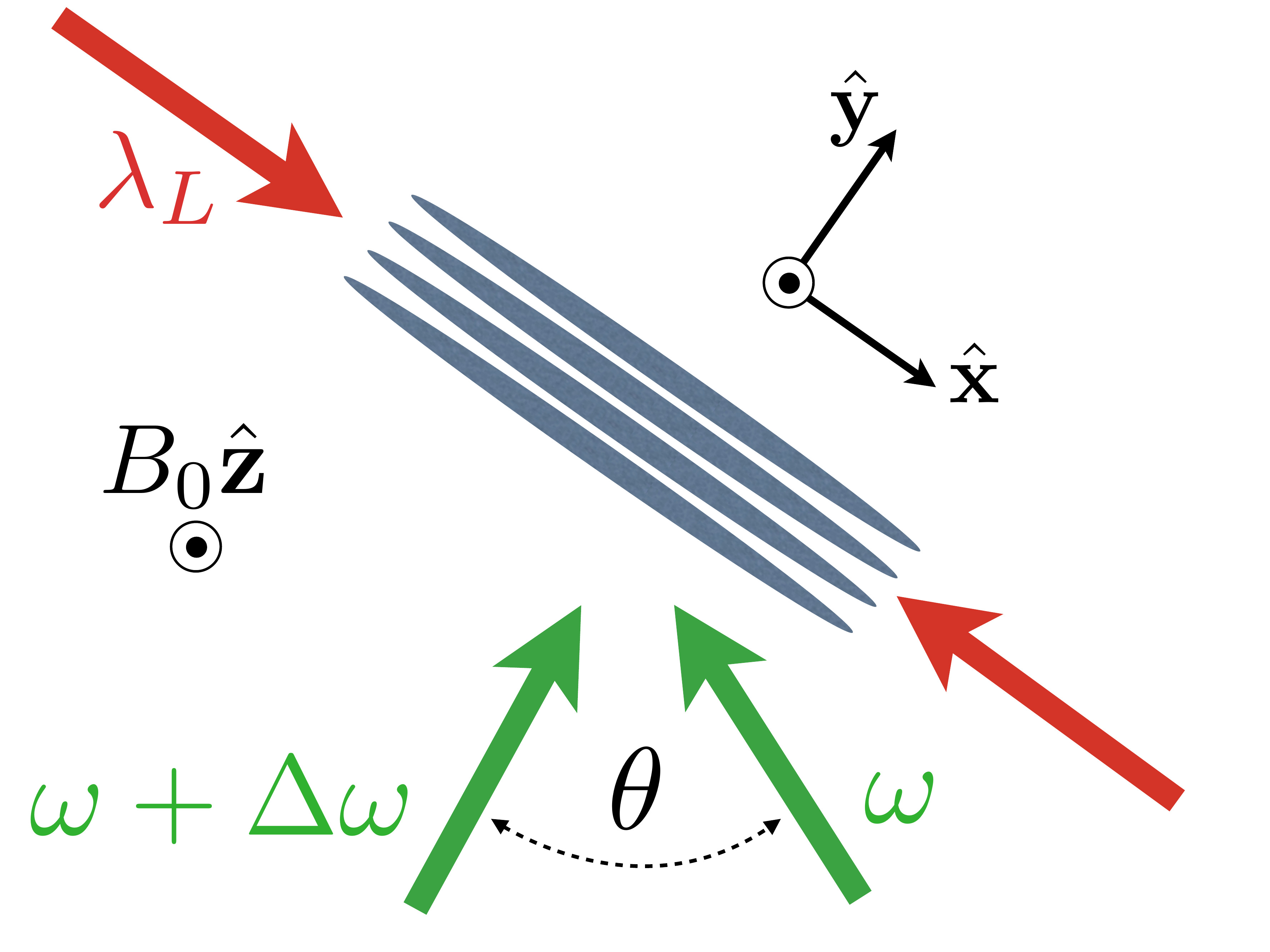}
\caption{Optical setup for the realization of the hybrid real+synthetic lattice. Red arrows: optical lattice beams. Green arrows: Raman beams.}
\label{fig:S1}
\end{figure}

\begin{figure}[b!]
\centering
\includegraphics[width=0.75\columnwidth]{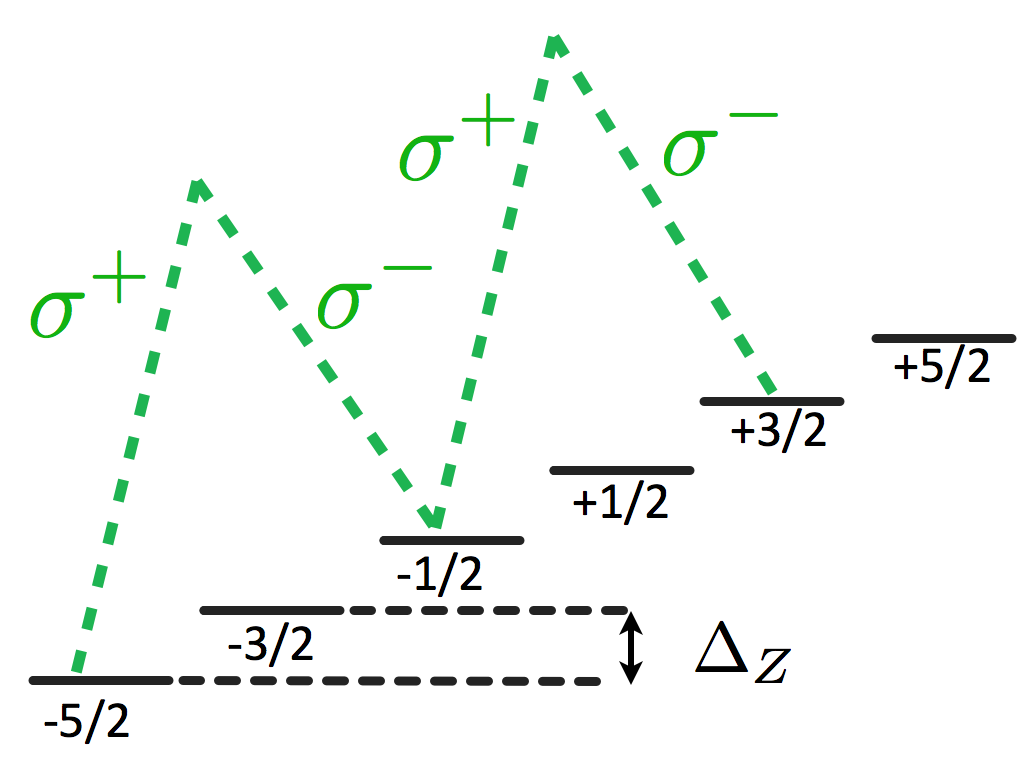}
\caption{Level diagram for the $\Delta m=\pm2$ Raman transitions providing the tunneling mechanism in the synthetic dimension of the Hall ribbon.}
\label{fig:S2}
\end{figure}

In this section we describe the laser configuration that produces a synthetic magnetic field in the Hall ribbon. The complex tunneling along the synthetic dimension $\hat{m}$ is implemented with two off-resonant $\lambda=556~\mbox{nm}$ laser beams with angular frequencies $\omega$ and $\omega + \Delta \omega$. We choose a detuning of $+1.87~\mbox{GHz}$ with respect to the narrow intercombination transition ${^1S_0}\rightarrow {^3P}_1(F'=7/2)$ in order to reduce the inelastic photon scattering rate.  The different periodicity of the Raman coupling $d_R=\lambda/\left[ 2\sin(\theta/2)\right]$, with $\theta$ being the relative angle between the two beams, with respect to the lattice spacing $d$, gives rise to a non-zero Peierls phase $\varphi=2 \pi (d/d_R)$. In our setup we choose $\theta=19.5^\circ$ that, considering the projection along the longitudinal axis of the fermionic chains, leads to a flux $\varphi=0.37\, \pi$ (see Fig. \ref{fig:S1}). The sign of the flux can be reversed by swapping the frequencies of the two Raman beams.
The Raman beams couple up to three nuclear spin projections of the $^1S_0$ ground states $m=(-5/2,-1/2,+3/2)$ via $\Delta m=\pm2$ processes. These states play the role of lattice sites in the extra-dimension $\hat{m}$, while the Rabi frequencies of the two-photon processes are equivalent to the intersite tunneling amplitudes.
The atoms are subjected to a $B_0=152$ Gauss  (real) magnetic field along $\hat{z}$ generating a linear Zeeman splitting $\Delta_Z=31.6(7)~\mbox{kHz}$ between adjacent nuclear spin components. In order to excite $\sigma^+/\sigma^-$ Raman transitions (Fig. \ref{fig:S2}), the beams frequency difference is set to $\Delta\omega/2\pi\simeq2\Delta_Z$ and the polarization is carefully chosen depending on the ladder configuration (see below).

The Raman Rabi frequencies and the light shifts are determined by considering the weighted sum of the matrix elements over the hyperfine structure $F'=(7/2,5/2,3/2)$ of the $^3P_1$ electronic excited state:\\
\begin{eqnarray}
\label{eq:}
\xi_\alpha &=& \sum_q\sum_{F'} \frac{|\Omega_{\alpha 1}^{q}(F')|^2+|\Omega_{\alpha 2}^{q}(F')|^2}{2\Delta_{F'}},  \\
\Omega_\alpha&=&\sum_{F'} \frac{\Omega^{+}_{\alpha 1}(F') \Omega^{-}_{\alpha 2}(F') }{2 \Delta_{F'}},          
\end{eqnarray}
where $q=(+, -, 0)$ is the photon polarization index, $\Delta_{F'}$ is the single-photon detuning with respect to the $F'$ state of the $^3P_1$ manifold and $\Omega^q_{\alpha i}$ is the single-photon Rabi frequency related to the $i=1,2$ Raman beam and $q$-polarization component. Here $\alpha$ is the synthetic lattice site index (see also Fig. 1 of the main text) and $\Omega_\alpha$ indicates the coupling between $\alpha$ and $\alpha+1$. With these definitions and considering also the overall harmonic confinement along $\hat{x}$, the total Hamiltonian of the system can be written as:
\begin{equation}
\label{eq:Hamiltonian}
\begin{split} H =&  \sum_{j=1}^N \sum_{\alpha=1}^3 \left[-t(c^\dagger_{j,\alpha}c_{j+1,\alpha} + h.c.) \right] \\&+ \sum_{j=1}^N \sum_{\alpha=1}^{2}\frac{\Omega_\alpha}{2}  (e^{i \varphi j } c^\dagger_{j, \alpha}c_{j, \alpha+1} + h.c.) \\&+\sum_{j=1}^N \sum_{\alpha=1}^3 \xi_\alpha n_{j,\alpha}+ \frac{W_x}{N^2} \sum_{j=1}^N \sum_{\alpha=1}^3  n_{j,\alpha} \left(j-\frac{N+1}{2} \right)^2.\end{split}
\end{equation}
In order to take into account the modification of the atomic density caused by the interparticle repulsion, we consider a slightly lower effective trap strength, $W_x \simeq 0.025 N^2 t$ (corresponding to a trap frequency $\sim40$ Hz along $\hat{x}$). Actually, the results depend only weakly on the particular trap frequency used in the calculations.

In order to calibrate the parameters of the system, we set the optical lattice beam along $\hat{x}$ at $V_{0x}=30E_R$ so that the coherent dynamics occurs only along the synthetic dimension ($t\ll\Omega_\alpha$), where we observe coherent spin oscillations up to 10 periods. Under these conditions the kinetic energy and harmonic confinement can be neglected and the Hamiltonian can be written as a $3\times3$ matrix in a frame rotating at frequency $2\Delta_Z$: 
\begin{equation}
\label{eq:}
\hat{H}/\hbar=\begin{pmatrix}[1.2] 
\xi_{{1}}-\delta&\Omega_{1}/2 & 0 \\
\Omega_{1}/2  & \xi_{2} & \Omega_{2}/2\\
0  & \Omega_{2}/2 & \xi_{3} + \delta
\end{pmatrix},
\end{equation}
where $\delta$ is the detuning with respect to the two-photon resonance and $\Omega_2=1.41\Omega_1$. 
By tuning the polarization of the two Raman beams we can choose whether to implement the 2- or the 3-leg Hall ribbon.

\vspace{3mm}
{\bf 2-leg ladder.} In this case, we choose $\hat{\bm {\varepsilon}} =(\hat{\varepsilon}_{+}+\hat{\varepsilon}_{-})/\sqrt{2}$, i.e. linear polarization in the plane $\hat{x}\hat{y}$, for both Raman beams. By setting $\delta=\xi_{1}-\xi_{2}$, we let the first two states to be resonantly coupled and exploit the light shift to isolate the third one, obtaining diagonal matrix elements $(\xi_{1}-\delta,\xi_{2},\xi_{3}+\delta)=(0,0,2.64)\Omega_{1}$ in the rotating frame . Under this condition the population of the third state is at most a few percent, and can thus be neglected.

\vspace{3mm}
{\bf 3-leg ladder.}  In order to resonantly couple all three states, we use an out-of-plane linear polarization $\hat{\bm {\varepsilon}} =(\hat{\varepsilon}_{+}+\hat{\varepsilon}_{-}+\hat{\varepsilon}_0)/\sqrt{3}$. In this case the light shifts are approximately the same for all the states (apart from corrections induced by the Zeeman shift of the excited state $^3P_1$) and by setting $\delta=\xi_{1}-\xi_{2}$  we obtain diagonal matrix elements $(\xi_{1}-\delta,\xi_{2},\xi_{3}+\delta)=(0,0,0.16)\Omega_{1}$. Under this condition the three states are all substantially populated.

\section{Adiabatic loading procedure}

\begin{figure}[t!]
\centering
\includegraphics[width=\columnwidth]{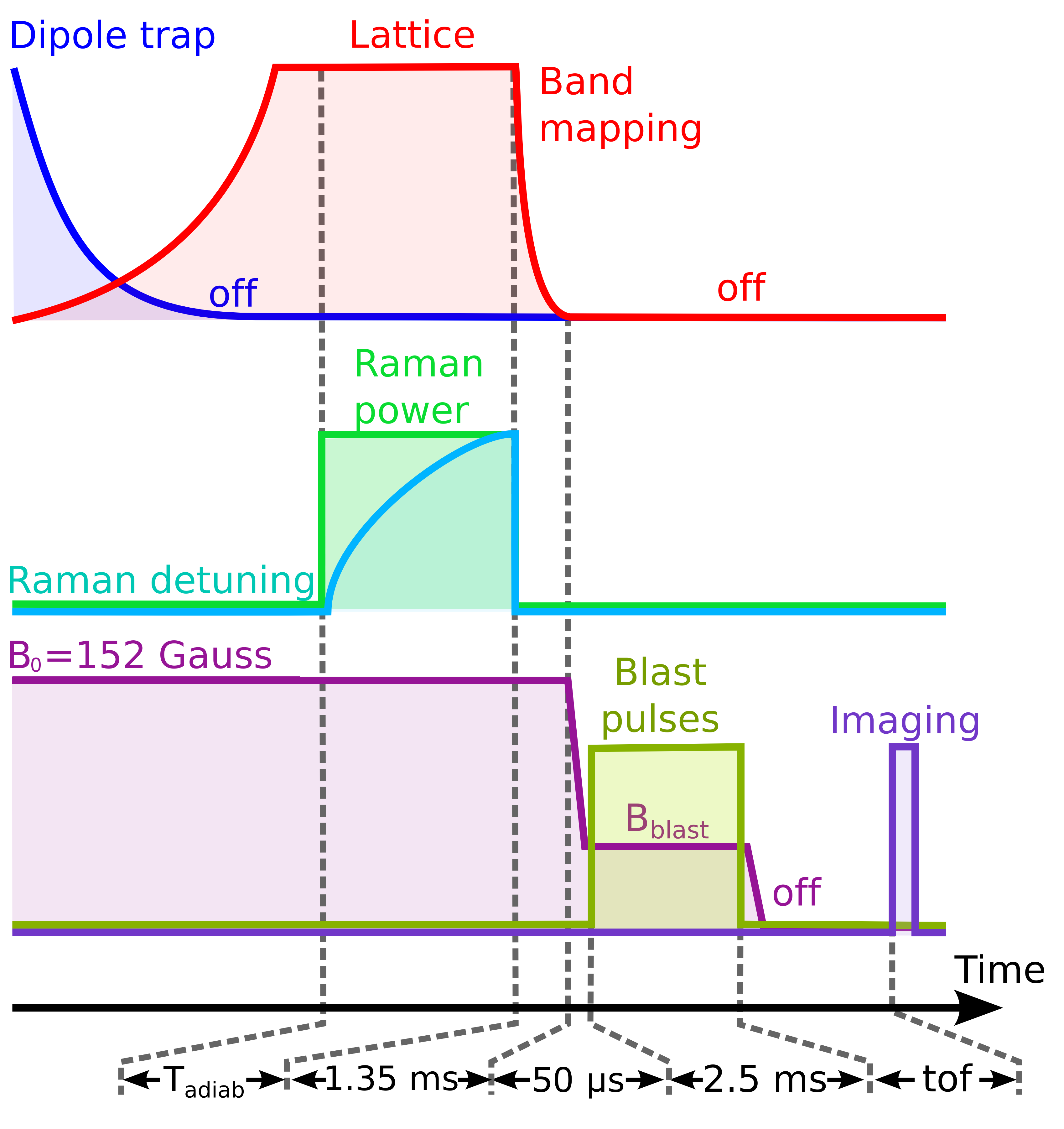}
\caption{Timing of the experimental procedure (see text for details).}
\label{fig:S3}
\end{figure}

In order to adiabatically load the equilibrium state of the Hall ribbon we start with a spin-polarized $m=-5/2$ Fermi gas and load it into the optical lattice (Fig. \ref{fig:S3}) with a 150 ms exponential ramp to the final values $V_{0x}=6.5 E_R$ and $V_{0y}=V_{0z}=30 E_R$.
During the lattice loading, the dipole trap is switched off with a 40 ms exponential ramp to decompress the atomic cloud and reduce the overall harmonic confinement to $\omega_x /2\pi \simeq 55 ~\mbox{Hz}$. After 5 ms, we switch on the Raman beams with an initial detuning $\delta_{in} \sim -25 \Omega_1$ and perform an exponential frequency sweep of the form:
\begin{equation}
\label{delta_t}
\delta(t)=\delta_{in} + (\delta_f - \delta_{in}) \left( \frac{1-e^{-t/\tau}}{1-e^{-T_{adiab}/\tau}}\right),
\end{equation}
where $\delta_f=\xi_1-\xi_2$.
The ramp duration $T_{adiab}$ ranges from 20 to 80 ms depending on the experimental configuration, with $\tau$ ranging from 5 to 25 ms accordingly. The adiabaticity of the whole process is verified experimentally by reversing the whole procedure to recover a spin-polarized Fermi gas.

\begin{figure}[t!]
\begin{center} 
\includegraphics[width=\columnwidth]{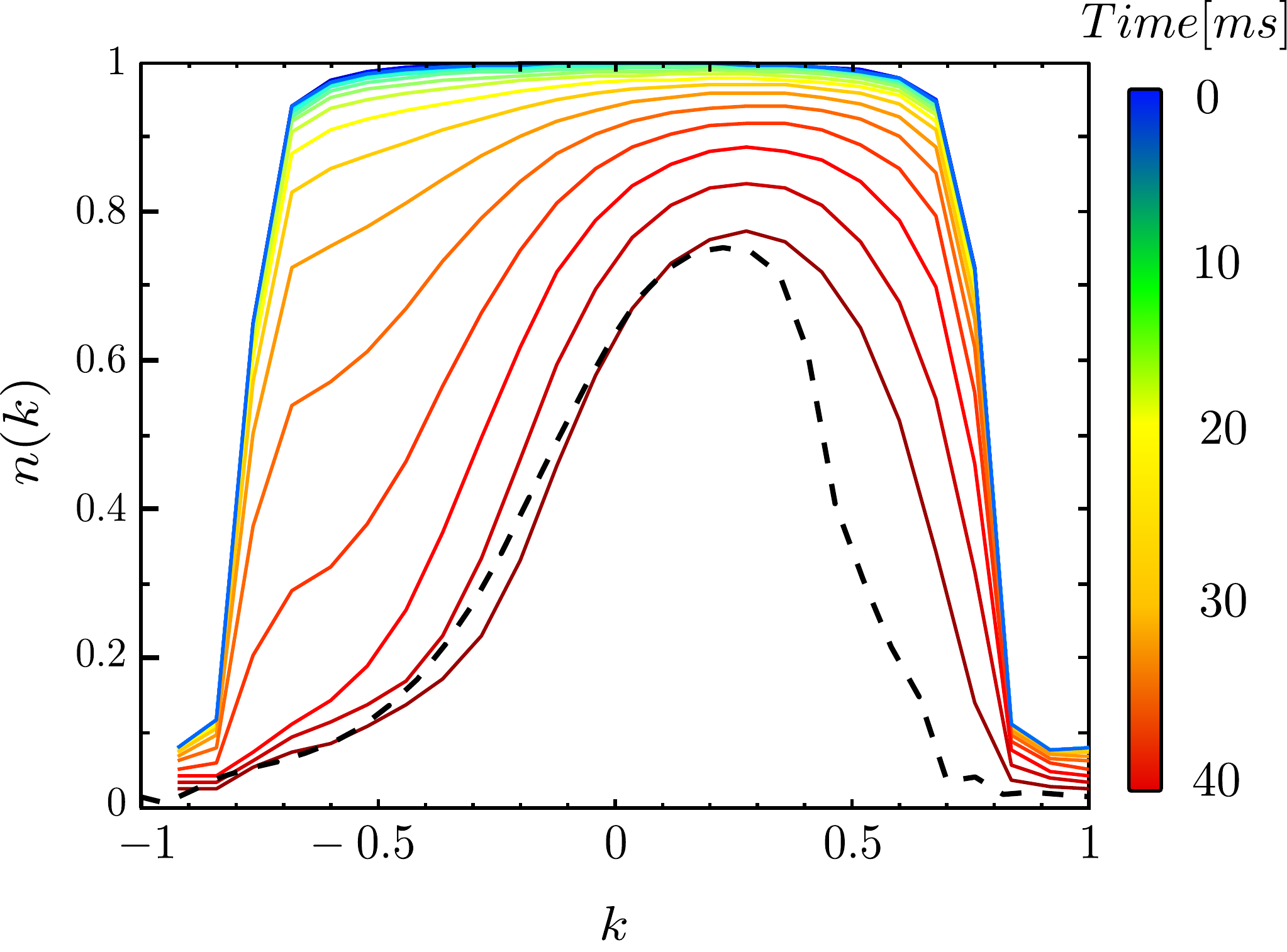}
\end{center}
\caption{Simulation of the adiabatic loading of the Hall ribbon. The solid lines show the calculated lattice momentum distribution for the $m=-5/2$ leg in a three-leg system at different times while the ramp of Eq. (\ref{delta_t}) is executed. The parameters for this evolution are $T_{adiab}=40$ ms and $\tau=12.5$ ms, while $\Omega_1/t = 3.65$. The dashed line shows the lattice momentum distribution as calculated from the ground state of the Hamiltonian with $\delta=\delta_f$.}
\label{adiabaticNK}
\end{figure}

In order to assess the validity of our adiabatic loading procedure we also conduct a numerical study.
We define a time-dependent Hamiltonian where the detuning plays the role of a time-dependent chemical potential. Then, using the same exponential detuning ramp used in the experimental setup, we simulate the evolution of the system starting from a spin-polarized Fermi gas in a one-dimensional optical lattice. We therefore compare the lattice momentum distribution calculated at the end of the evolution with the one obtained from the ground state of the Hamiltonian with $\delta=\delta_f$ (see Fig. \ref{adiabaticNK}). The comparison between the two distributions indicates that the ramp in eq. (\ref{delta_t}) creates only a small number of excitations in the preparation of the Hall ribbon.

\section{Spin-selective imaging}

After the ground state loading, we suddenly switch off the Raman coupling, therefore freezing the population along the synthetic direction. Then we switch off the real lattice with a $t_{map}=1.35$ ms exponential ramp, which is slower than the timescale associated 
to the lattice band-gap. This procedure allows us to adiabatically map the lattice momentum distribution along the chains onto the atomic velocity distribution \cite{greiner2001,kohl2005}, which is measured by absorption imaging after 23 ms of ballistic expansion.

In order to perform a single-site imaging along the synthetic direction, we use a sequence of spin-selective laser pulses (``blast'' pulses), in resonance with different components of the narrow intercombination transition $^1S_0\rightarrow{^3}P_1 (F'=7/2)$, to remove atoms in all the spin states but one. The sequence is carried out during the first 2.5 ms of ballistic expansion. At this time the (real) magnetic field is $B_{blast}=15$ Gauss (see Fig. \ref{fig:S3}), leading to a Zeeman shift $\Delta_Z\sim 50 (\Gamma/2\pi)$ between adjacent spin components in the $^3P_1$ manifold (where $\Gamma=2\pi \times 180$ kHz is the natural linewidth of the transition). This  separation allows us to use two opportunely detuned $\sigma^+$ and $\sigma^-$ beams to remove the unwanted spin population, without causing any heating to the selected spin state left in the expanding cloud. After ballistic expansion, absorption imaging is performed on the dipole-allowed $^1S_0\rightarrow{^1}P_1\,(F'=7/2)$ transition at 399 nm (with natural linewidth $\Gamma'=2\pi\times 28.9~\mbox{MHz}$) with a resonant pulse along $\hat{z}$.

\section{Image analysis}

{\bf Lattice momentum distribution and $J$ vs $\Omega/t$.}
The lattice momentum unbalance in a given leg is measured by taking the average of at least 30 images for each spin state with the procedure shown in Fig. \ref{fig:S3}. 
The geometric center $k_0$ of the lattice momentum distribution $n(k)$ is evaluated by acquiring images with Raman beams out of resonance, namely with tunneling occurring only along the real direction. Indeed, in this configuration no chirality is induced in the system, which is purely one-dimensional resulting in a momentum distribution symmetric around $k=0$. All the images are  integrated along the $\hat{y}$ direction and shifted by $k_0$ in order to obtain the $n(k)$ curves for the different legs.
In order to remove residual gradients or fringes due to imperfections in the imaging setup, also background images are acquired, averaged and subtracted from $n(k)$. Finally, we normalize $n(k)$ for each spin state in such a way as to have $\int n(k) dk=1$ (here, and in the figures of the main text, $k$ is expressed in units of the lattice wavenumber $k_L=\pi/d$, so that $k=\pm1$ correspond to the boundaries of the first Brillouin zone). The quantity $J=\int_0^1 h(k)dk$ for a given spin state is calculated by analyzing the average image, while the errors are the standard deviation of the statistical distribution obtained after performing bootstrapping on the experimental data.

\vspace{3mm}

{\bf Edge-cyclotron orbits.} In the quench dynamics experiments shown in Fig. 4, in order to study the evolution along real space, we determine the average position $\langle x \rangle$ of the atomic cloud assuming the validity of the semiclassical equation of motion along the real lattice \cite{note2}. In this framework, considering the lowest band dispersion as $\mathcal{E}(k)=2 t \left[1- \cos(k d)\right]$, the velocity of the $k$-component of the Fermi sea is
\begin{equation}
\label{eq:}
v_k=\frac{1}{\hbar}\frac{\partial \mathcal{E}(k)}{\partial k}=\frac{2 t d}{\hbar} \sin(k d).
\end{equation}
Then, knowing the tunneling $t$, we can measure the average velocity of the whole cloud at a given time $\tau$ from the experimental lattice momentum distribution $n(k,\tau)$:
\begin{equation}
\label{eq:}
\langle v(\tau)  \rangle=\frac{2 t d}{\hbar}  {\displaystyle \int  n(k,\tau) \sin(k d)dk}   .
\end{equation}
Finally, to obtain the average position at a given time $\tau$, we integrate the average velocity over time:
\begin{equation}
\label{eq:}
\langle x(\tau)  \rangle = {\displaystyle \int_0^\tau \langle v (\tau')\rangle}d\tau'.
\end{equation}
In order to estimate the errors bars of the edge-cyclotron orbits in Fig. 4d of the main text, we perform a bootstrap analysis over a set of different $\langle x(\tau)\rangle$ reconstructed from a random sampling of the experimental images.

\section{Theoretical description}

In this section, we provide more details on the simulations presented in the paper. 

\vspace{3mm}

{\bf $J$ versus $\Omega/t$.} In order to quantify the chiral character of both 2- and 3-leg ladders, we defined the quantity
\begin{equation}
J = \int_0^1 h(k) dk 
\end{equation}
with
\begin{equation}
h(k)=n(k)-n(-k). 
\end{equation}
We computed this quantity for the Hamiltonian in Eq.~(\ref{eq:Hamiltonian}) in the central region of the system, with a peak density of ~0.8/0.9 particles per rung in the center, and an approximate size of up to 15 $\mu$m (corresponding to up to $L=40$ lattice rungs with density  $>10^{-5}$). For the data in Figs. 2 and 3 of the main text, we considered a single Hall ribbon comprising 75 sites and 20 particles, and had the light shifts as the only free parameters in the microscopic Hamiltonian for the 3-leg case (in the 2-leg case the results are not very sensitive to the light shift on the $-1/2$ leg, so that we kept the one extracted from the calibration, as described in Section S.II). In the 3-leg case, we found that the best quantitative agreement with the experimental data is obtained by using a set of light shifts ${\xi_\alpha}$ that slightly differ from the nominal ones. In particular, for the simulations displayed in Fig. 3c, we used  $(\xi_{1},\xi_{2},\xi_{3})=(0,0.1,-0.26)\Omega_{1}$. The discrepancy between these best-fitting values and the nominal ones can be easily ascribed to the experimental uncertainty in the Raman beams polarization, whose variations have a very strong effect on the light shifts.

In order to take into account thermal fluctuations and possible effects of the non-adiabatic loading during the lattice ramp, we simulated systems with an average of 20\% thermal excitations above the Fermi sea (both the 2-leg and 3-leg case), and performed averages over up to 200 configurations. These results are approximately recovered by investigating a finite-temperature partition function with temperature $T\simeq 0.4 t$. The width of the theoretical curves in Figs. 2d and 3c, representing the effect of thermal excitations, has been evaluated as the 68.27\% percentile interval centered around the mean value of the statistical distribution obtained with a bootstrapping analysis of 100 realizations of the simulation.

In addition, we have performed simulations over the entire system (up to $2\times10^4$ particles in $\sim 1000$ independent ribbons) using a local density approximation in the grand canonical ensemble. The results (not displayed) are qualitatively the same as for the simple model used here.

\vspace{3mm}

{\bf Chiral dynamics at the edge.} For the quench dynamics illustrated in Fig. 4 of the manuscript, we solved numerically the time-evolution of systems at different densities. Indeed, it turned out that assuming a single value for the density is usually not sufficient to capture the full time-dependent dynamics at the edge. As a matter of fact, in the different ladders of the system, each with a different density, there is a very different fraction of particles participating in the edge dynamics. Whilst not affecting drastically the behavior of $J$, this inhomogeneity has strong effects on the single edge-dynamics probed by a quench. In order to take into account the inhomogeneity effect, we have simulated systems at various system sizes up to $L=55$ lattice rungs and different densities. For the data shown in Fig. 4, we considered a system of $L=35$ rungs and averaged over a set of different densities $(0.2,0.4,0.6,0.8)$. We notice that, when taken singularly, none of these density realizations captures the system dynamics correctly, while the average gives a very good agreement with the experimental results.


\begin{thebibliography}{99}
\bibitem[16]{note1}
During the completion of this manuscript we became aware of a closely related experimental work with Raman-coupled Bose-Einstein condensates [26].
\end{thebibliography}
\begin{thebibliography}{99}
\bibitem[29]{note2}
Notice that, for the full two-dimensional case, another term, taking into account the Berry phase in momentum space, is also affecting $\langle x\rangle$ at the level of semi-classical trajectories (see, e.g., Ref. [30]). This is however not applicable for the present case, where the real-space dynamics of the particles is strictly one-dimensional.
\end{thebibliography}
\end{document}